\documentclass[showpacs,showkeys,preprintnumbers,ansmath,amssymb]{revtex4}
\usepackage[dvips]{graphicx}
\newcommand{\be}{\begin{equation}}
\newcommand{\ee}{\end{equation}}
\newcommand{\bea}{\begin{eqnarray}}
\newcommand{\eea}{\end{eqnarray}}
\newcommand{\bm}{\begin{mathletters}}
\newcommand{\eml}{\end{mathletters}}

\begin{document}

\title{Synthesis of superheavy elements beyond Z=118}

\author{M. Mirea$^{1}$, D.S. Delion$^{1,2}$ and A. S\u andulescu$^{2,3}$}
\affiliation{
$^{1}$National Institute of Physics and Nuclear Engineering,\\
407 Atomi\c stilor, Bucharest-M\u agurele, Romania \\
$^{2}$Academy of Romanian Scientists \\
Splaiul Independen\c tei 54, Bucharest, Romania \\
$^{3}$Institute for Advanced Studies in Physics,\\
Calea Victoriei 129, Bucharest, Romania}

\begin{abstract}
{We investigate the cold fission/fusion paths of superheavy nuclei within
the two center shell model, in order to find the best projectile-target
combinations of their production.
The fission/fusion yields are estimated by using the semiclassical approach.
We predict several asymmetric combinations of relative long living fragments,
which can be used in fusion experiments of superheavy nuclei with $Z>118$.}
\end{abstract}

\vskip1cm

\pacs{21.10.Dr, 21.10.Tg, 25.70.Jj, 25.85.Ca}

\keywords{Cold fission, Superheavy nuclei, Potential surface, Cold valleys,
Two center shell model, Woods-Saxon potential}

\maketitle
%\date{\today}

%\newpage

\rm

\setcounter{equation}{0}
\renewcommand{\theequation}{\arabic{equation}}

The synthesis of superheavy elements beyond $Z=104$, suggested by Flerov
\cite{Fle69}, was predicted within the so-called fragmentation theory in
Ref. \cite{San76} by using the cold valleys in the potential energy surface
between different combinations, giving the same compound nucleus.
Soon it was shown in Refs. \cite{Gup77,Gup77a} that the most favorable
combinations with $Z\geq 104$ are connected with the so-called Pb potential
valley, i.e. the same valley of the heavy cluster emission \cite{San80}.

Due to the double magicity of $^{48}$Ca, similar with $^{208}$Pb, in Ref.
\cite{Gup77a} it was proposed $^{48}$Ca as a projectile on various
transuranium targets.
Indeed, the production of many superheavy elements with $Z\leq 118$
(corresponding to the last stable element Cf) during last three decades
was mainly based on this idea \cite{Oga01,Oga04,Hof00,Hof04}.

The formation of superheavy compound systems by fusion was intensivelly
explored \cite{Den00,Smo01,Zag01}. On the other side the investigation
of experimental data concerning fusion and fission of superheavy nuclei
with $Z=112,114,116$, together with data on survival probability of these
nuclei in evaporation channels with 3-4 neutrons, revealed the fact
that the fission barriers are quite high, leading to a relative
high stability of such systems \cite{Itk02}.

The main tool to investigate such nuclei is almost exclusively based upon
the investigation of $\alpha$-decay chains.
In the last decade several papers were devoted to the calculation
of $\alpha$-decay half-lives in this region \cite{Gam05}.
All these approaches can be considered as phenomenological ones,
based essentially on the Gamow $\alpha$-nucleus potential picture \cite{Gam28}.
The recent microscopic estimate of the $\alpha$-particle preformation factor,
by using shell model single particle orbitals, performed in Ref. \cite{Del04},
showed that the strong change of the $Q$-value along neutron chains can
by explained only by supposing the existence of an $\alpha$-cluster component
in heavy and superheavy emitters.

The $\alpha$-particle emission is connected with the "lightest"
side of the cold valley on the fragmentation potential surface.
On the other hand the "heaviest" side of the cold valley
is given by the cold fission process, i.e. the emission of
two fragments with similar masses in their ground states.
Between theses limits there is a broad region of cold  heavy cluster decays.
The aim of this work is to evidence cold fission valleys, which can be
good candidates for the production of superheavy elements with $Z\geq 118$,
by using the inverse, fusion process. We extend the analysis performed
in Ref. \cite{Del07} within a simple model, to a more
reliable microscopic approach to estimate the fission/fusion barrier,
given by a new version of the
 Super Asymmetric Two Center Shell Model \cite{Mir98}. This version
solves a Woods-Saxon potential \cite{Mir07} in terms of the two center 
prescriptions 
and provides two additional degrees of freedom, that is, the deformations
of the fragments.
The deformation energy of the di-nuclear system is the sum between the
liquid drop energy and the shells effects, including pairing corrections.
The macroscopic energy is obtained within the framework of the 
Yukawa-plus-exponential model extended to binary systems with different
charge densities \cite{Poe80}.
Strutinsky correction prescriptions \cite{Bra72} were computed on the basis of
a new version of the two center model with Woods-Saxon deformed potentials
of fragments. We considered only cold fission/fusion process. Consequently
the deformations of the initial and final nuclei are given by their ground
state values of Ref. \cite{Mol95}. 

The penetrability, coresponding to some binary partition, defines the
isotopic yield and it is characterized by the difference between the
nuclear plus Coulomb potential and the $Q$-value.
For a given initial nucleus $(Z,A)$ this quantity, called drivind potential,
depends upon the charge, mass numbers of a given fragment (we will consider
the second one) and the inter-fragment distance.
For a fixed combination $A=A_1+A_2$ the driving potential has a minimum at
the charge equilibration point $Z_2$, which we will not mention in the
following, i.e.
\bea
\label{poten}
V(A_2,R)=V_N(A_2,R)+V_C(A_2,R+B(Z_1,A_1)+B(Z_2,A_2)~,
\eea
where $V_N(A_2,R)$ is the nuclear and $V_C(A_2,R)$ Coulomb inter-fragment
potential. Here the binding energy of the initial nucleus is not considered,
because it has the same value for all binary partitions.
For deformed nuclei, due to the fact that the largest emission probability
corresponds to the lowest barrier, this potential decreases in the direction
of the largest fragment radius.
We mention that a very convincing theoretical evidence that the cold fission
has a sub-barrier character and depends upon deformation parameters was the
calculation of penetration factors, using a double the double folding
inter-fragment potential, with M3Y plus Coulomb nucleon-nucleon forces.
This simple estimate was able to reproduce the gross features of the binary
cold fragmentation isotopic yields from $^{252}$Cf \cite{San98}.
Here it was shown the important role not only of quadrupole, but also of
hexadecapole deformation parameters on emission probability.

\begin{figure}[ht]
\begin{center}
\includegraphics[width=9cm]{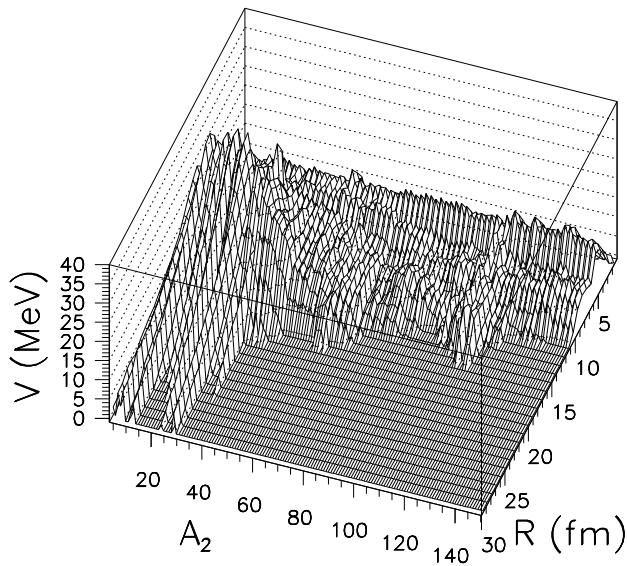}
\caption{\it
The two dimentional potential $V(A_2,R)$ versus the fragment mass number
and the inter-fragment distance $R$,
for the hypothetical compound nucleus $^{300}_{120}X$.}
\label{fig1}
\end{center}
\end{figure}

Our first step is to estimate the driving potential (\ref{poten}) for nuclei
beyond Z=118.
First of all we have checked the validity of the two center shell model,
by reproducing qualitatively cold isotopic yields from $^{252}$Cf \cite{San98}.
Then, we computed the driving potential for the hypothetical compound
nucleus $^{300}_{120}X$. This nucleus has a larger amount of neutrons
than the measured superheavy combinations with Z=116 and 118 \cite{Oga04a}.
In Fig. \ref{fig1} we plotted the two dimensional potential surface
$V(A_2,R)$ versus the fragment mass number of the second partner and the
inter-fragment distance $R$.
It is clear that the cold fission/fusion process is very much hindered in
the region of the large maximum between $A_2=4$ and $A_2=40$. 

\begin{figure}[ht]
\begin{center}
\includegraphics[width=9cm]{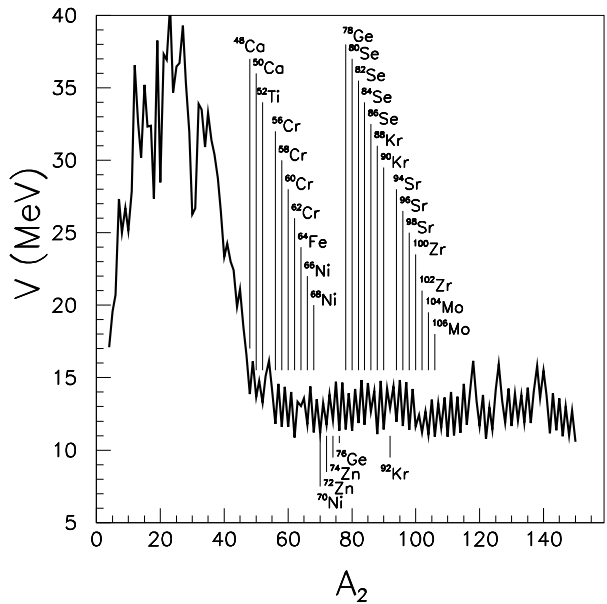}
\caption{\it
The maximum value of the potential surface in Fig. \ref{fig1} with respect
to the inter-fragment radius $R$,  versus the fragment mass number.}
\label{fig2}
\end{center}
\end{figure}

This is especialy clear from Fig. \ref{fig2}, where we plotted the maximum
value of the potential surface in Fig. \ref{fig1} with respect to
the inter-fragment radius $R$, as a function of the fragment mass number.

The first relevant minimum of the potential surface in the region $A_2>40$
corresponds to the already mentioned double magic nucleus $^{48}$Ca,
but unfortunately its binary partner $^{252}$Fm is unstable, as can be
seen from Table 1.

\begin{figure}[ht]
\begin{center}
\includegraphics[width=9cm]{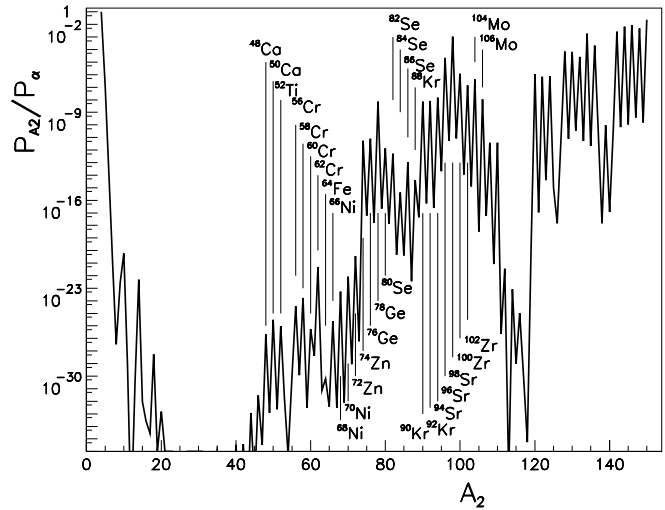}
\caption{\it
The penetrabilities corresponding to the potential surface in Fig. \ref{fig1}
versus the mass number of the first fragment.}
\label{fig3}
\end{center}
\end{figure}

In order to search for reliable binary candidates, producing the above
mentioned hypothetical compound nucleus, it is necessary to investigate
the penetration factor. This quantity can be estimated,
as usually, by using the semiclassical integral
\bea
\label{penetr}
P_{A_2}=exp\left\lbrace-2\int_{R_1}^{R_2}
\sqrt{\frac{2\mu}{\hbar^2}\left[V(A_2,R)-B(Z,A)\right]} dR\right\rbrace~,
\eea
between internal and external turning points.
In Fig. \ref{fig3} is given the penetrability ratio $R=P_{A_2}/P_{\alpha}$,
corresponding to the potential surface in Fig. \ref{fig1}.
Several maxima, comparable or larger than the mentioned combination
$^{48}$Ca+$^{252}$Fm, are present in this figure.
They correspond to Cr, Fe, Ni, Zn, Ge, Se projectiles. The most promising
are the projectiles beyond Zn isotopes, where the penetrability increases by
nine orders of magnitude from $^{72}$Zn to $^{74}$Zn, as can also be seen from
Table 1. The region of maximal values corresponds to Sr+Pb combinations,
i.e. to the already mentioned Pb valley.

On the other hand the target-projectile combinations should be relative stable.
In Table 1 are given the fragment half-lives of these binary combinations,
with $A_2>40$, corresponding to local maxima of the penetrability.
Unfortunately the half-lives of $^{72,74}$Zn isoptopes and their partners
are fast decreasing. 

\newpage
\begin{center}
{\it Table 1 \\ Mass numbers and half-lives of binary partners
for the hypothetical compound nucleus $^{300}_{120}X$.
In the last column is given the penetrability ratio 
$R=P_{A_2}/P_{\alpha}$.}
\vskip3mm
\begin{tabular}{|l|r|l|r|r|r|}
\hline
$^{A_1}_{Z_1}X_1$ & $T_1~~~$ & $^{A_2}_{Z_2}X_2$ & $T_2~~~$ & $R$~~~~~~ \cr
\hline
$^{48}_{20}$Ca & $10^{19}$ y & $^{252}_{100}$Fm & 25.4 h      & 2.1 10$^{-26}$  \cr
$^{50}_{20}$Ca & 13.9 s      & $^{250}_{100}$Fm & 30 m        & 3.1 10$^{-25}$  \cr
$^{52}_{22}$Ti & 1.7 s       & $^{248}_{98}$Cf & 333.5 d      & 9.5 10$^{-26}$  \cr
$^{56}_{24}$Cr & 5.94 m      & $^{244}_{96}$Cm & 18.1 y       & 3.7 10$^{-24}$  \cr
$^{58}_{24}$Cr & 7 s         & $^{242}_{96}$Cm & 162.8 d      & 1.7 10$^{-23}$  \cr
$^{60}_{24}$Cr & 490 ms      & $^{240}_{96}$Cm & 27 d         & 5.5 10$^{-26}$  \cr
$^{62}_{24}$Cr & 160 ms      & $^{238}_{96}$Cm & 2.4 h        & 4.8 10$^{-21}$  \cr
$^{64}_{26}$Fe & 2 s         & $^{236}_{94}$Pu & 2.9 y        & 6.3 10$^{-30}$  \cr
$^{66}_{28}$Ni & 54 h        & $^{234}_{92}$U  & 2.5 $10^5$ y & 2.5 10$^{-25}$  \cr
$^{68}_{28}$Ni & 29 s        & $^{232}_{92}$U  & 68.9 y       & 5.4 10$^{-23}$  \cr
$^{70}_{28}$Ni & 1 $\mu$s    & $^{230}_{92}$U  & 20.8 d       & 8.1 10$^{-22}$  \cr
$^{72}_{30}$Zn & 46.5 h      & $^{228}_{90}$Th & 1.9 y        & 3.7 10$^{-20}$  \cr
$^{74}_{30}$Zn & 95.6 s      & $^{226}_{90}$Th & 30.6 m       & 5.7 10$^{-11}$  \cr
\hline
\end{tabular}
\begin{tabular}{|l|r|l|r|r|r|}
\hline
$^{A_1}_{Z_1}X_1$ & $T_1~~~$ & $^{A_2}_{Z_2}X_2$ & $T_2~~~$ & $R$~~~~~~ \cr
\hline
$^{76}_{32}$Ge & $10^{25}$ y & $^{224}_{88}$Ra & 3.7 d        & 8.1 10$^{-11}$  \cr
$^{78}_{32}$Ge & 88 m        & $^{222}_{88}$Ra & 38.0 s       & 7.1 10$^{-08}$  \cr
$^{80}_{34}$Se & stable      & $^{220}_{86}$Rn & 55.6 s       & 1.4 10$^{-11}$  \cr
$^{82}_{34}$Se & $10^{19}$ y & $^{218}_{86}$Rn & 35 ms        & 5.2 10$^{-12}$  \cr
$^{84}_{34}$Se & 3.1 m       & $^{216}_{86}$Rn & 45 $\mu$s    & 4.1 10$^{-15}$  \cr
$^{86}_{34}$Se & 15.3 s      & $^{214}_{86}$Rn & 270 ns       & 1.0 10$^{-12}$  \cr
$^{88}_{36}$Kr & 2.84 h      & $^{212}_{84}$Po & 299 ns       & 3.9 10$^{-14}$  \cr
$^{90}_{36}$Kr & 32.3 s      & $^{210}_{84}$Po & 138.4 d      & 7.4 10$^{-08}$  \cr
$^{92}_{36}$Kr & 1.84 s      & $^{208}_{84}$Po & 2.9 y        & 8.1 10$^{-08}$  \cr
$^{94}_{38}$Sr & 75.3 s      & $^{206}_{82}$Pb & stable       & 1.4 10$^{-07}$  \cr
$^{96}_{38}$Sr & 1 s         & $^{204}_{82}$Pb & stable       & 2.2 10$^{-04}$  \cr
$^{98}_{38}$Sr & 653 ms      & $^{202}_{82}$Pb & 5.2 $10^3$ y & 1.1 10$^{-02}$  \cr
& & & & \cr
\hline
\end{tabular}
\end{center}

In order to compare the data in Table 1 with those corresponding to
some experimentally detected superheavy elements, in Table 2 we give 
penetrability ratios of three measured fusion reactions.

We also compared the penetrabilities in Table 1 with the corresponding
quantities of some close elements which have been experimentally reported
by using $^{48}$Ca as projectile.
The reactions given in the first and second lines of the Table 2
have penetrability ratios around $R\approx 10^{-24}$. 
The corresponding values of the first two lines in Table 1 are not
far from this number.

\begin{center}
{\it Table 2 \\ Mass numbers and half-lives of binary partners
for the experimentally detected compound nuclei $^{A}_{Z}X$.
In the last columns is given the penetrability ratio $R=P_{A_2}/P_{\alpha}$
and the quoted reference.}
\vskip3mm
\begin{tabular}{|l|r|l|r|r|r|r|c|}
\hline
$^{A_1}_{Z_1}X_1$ & $T_1~~~$ & $^{A_2}_{Z_2}X_2$ & $T_2~~~$ &
$^{A}_{Z}X$ & $R$~~~~~~ & Ref. \cr
\hline
$^{48}_{20}$Ca & $10^{19}$ y & $^{244}_{96}$Cm & 18.1 y & $^{292}_{116}$X & 1.7 10$^{-24}$ & \cite{Oga04a} \cr
$^{48}_{20}$Ca & $10^{19}$ y & $^{246}_{98}$Cf & 35.7 h & $^{294}_{118}$X & 1.1 10$^{-24}$ & \cite{Oga04a} \cr
$^{70}_{30}$Zn & stable      & $^{208}_{82}$Pb & stable & $^{278}_{112}$X & 5.0 10$^{-11}$ & \cite{Hof96} \cr
\hline
\end{tabular}
\end{center}

The penetrability ratio, corresponding to the third reaction 
$^{208}$Pb+$^{70}$Zn$\rightarrow^{278}_{112}$X \cite{Hof96},
is $R=P_{A_2}/P_{\alpha}\approx$ 5 $10^{-11}$.
Practically all combination beyond $^{74}$Zn, giving the hypothetical
element $^{300}_{120}$X, have the penetrability ratios
larger than this value. In Table 1 we remark that in the
combination $^{76}$Ge+$^{224}$Ra both partners have enough large
half-lives to be used in fusion experiments.

Concluding, we computed the potential energy surface for different binary
combinations, giving the superheavy compound nucleus $^{300}_{120}$X,
by using the Two Center Shell Model.
We evidenced binary partners whose penetration factors are comparable with
similar combinations, producing close superheavy elements which were
experimentally measured.
An extensive analysis on fission/fusion channels for all possible isotopes
in this region of superheavy elements, by using this performant version
of the Two Center Shell Model, is under way.

%\newpage

%\end{document}

\end{document}